\documentclass[aps,prc,superscriptaddress,showpacs,floatfix,nofootinbib,twocolumn]{revtex4-1}
\usepackage{amsmath,graphicx,float,hyperref}
\def\bra{\langle}
\def\ket{\rangle}
\newcommand{\trento}{T$\mathrel{\protect\raisebox{-2.1pt}{R}}$ENTo}

\begin{document}
\title{Relative flow fluctuations as a probe of initial state fluctuations}

\author{Giuliano Giacalone}
\affiliation{Institut de physique th\'eorique, Universit\'e Paris Saclay, CNRS,
CEA, F-91191 Gif-sur-Yvette, France}
\author{Jacquelyn Noronha-Hostler}
\affiliation{Department of Physics, University of Houston, Houston TX 77204, USA}
\author{Jean-Yves Ollitrault}
\affiliation{Institut de physique th\'eorique, Universit\'e Paris Saclay, CNRS, CEA, F-91191 Gif-sur-Yvette, France} 

\begin{abstract}
Elliptic flow, $v_2$, and triangular flow, $v_3$, are to a good approximation linearly proportional to the corresponding spatial anisotropies of the initial density profile, $\varepsilon_2$ and $\varepsilon_3$.
Using event-by-event hydrodynamic simulations, we point out when deviations from this linear scaling are to be expected.  
When these deviations are negligible, relative $v_n$ fluctuations are equal to relative $\varepsilon_n$ fluctuations, and one can directly probe models of initial conditions using ratios of cumulants, for instance $v_n\{4\}/v_n\{2\}$. 
We argue that existing models of initial conditions tend to overestimate flow fluctuations in central Pb+Pb collisions, and to underestimate them in peripheral collisions. 
We make predictions for $v_3\{6\}$ in noncentral Pb+Pb collisions, and for $v_3\{4\}$ and $v_3\{6\}$ in high-multiplicity p+Pb collisions.  
\end{abstract}

\maketitle

\section{introduction}
Anisotropic flow is the key observable providing evidence for the creation of a collective medium in ultrarelativistic heavy-ion collisions.
In the current paradigm of bulk particle production~\cite{Luzum:2011mm}, anisotropic flow emerges from the hydrodynamical response of the created medium to the anisotropies of its initial energy density profile~\cite{Teaney:2010vd}.
Hydrodynamic simulations~\cite{Gardim:2011xv,Niemi:2012aj,Gardim:2014tya} show that elliptic flow, $v_2$, and triangular flow, $v_3$, correlate almost linearly with the initial eccentricity, $\varepsilon_2$, and triangularity, $\varepsilon_3$, of the system.
Since the initial energy density profile is shaped out of stochastic nucleon-nucleon interactions, both initial anisotropies and flow coefficients fluctuate on a event-by-event basis~\cite{Alver:2006wh}.
To the extent that $v_n$ is proportional to  $\varepsilon_n$, the probability distribution of $v_n$~\cite{Aad:2013xma} coincides, up to a global rescaling, with the probability distribution of $\varepsilon_n$~\cite{Renk:2014jja,Yan:2014nsa}. 
The latter is provided by models of initial conditions.

Many models of initial conditions have been proposed for proton-nucleus and nucleus-nucleus collisions. 
Some are based on variations of the Glauber Monte Carlo model~\cite{Miller:2007ri,Alvioli:2009ab,Rybczynski:2013yba,Loizides:2016djv,Zakharov:2016gyu}, others are more directly inspired from high-energy QCD, and involve, in particular, the  idea of gluon saturation \cite{Drescher:2006ca,Albacete:2010ad,Werner:2010aa,Schenke:2012wb,Albacete:2014fwa,Niemi:2015qia}. 
The initial anisotropies $\varepsilon_n$ probe the geometrical shape of the initial density profile, and, thus, provide information which is independent of the final multiplicity distribution, which is the typical observable to which models are tuned. 
Therefore, observables which can be linked to initial anisotropies allow one to further constrain initial condition models, and to eventually  obtain new insight into the early dynamics of the collision. 

In this paper, we analyze the relative fluctuations of $v_2$ and $v_3$ in p+Pb and Pb+Pb collisions at CERN Large Hadron Collider (LHC) energies.
The observables we choose for this analysis are ratios of cumulants of the distribution of $v_n$, whose definition is recalled in Sec.~\ref{sec:2}.
In Sec.~\ref{sec:3}, we compute the lowest non-trivial ratios of cumulants, $v_2\{4\}/v_2\{2\}$ and $v_3\{4\}/v_3\{2\}$, in event-by-event hydrodynamic simulations of Pb+Pb collisions, and we determine in which centrality intervals they are compatible with the ratios of cumulants of the corresponding initial anisotropies, $\varepsilon_n$. 
In these centrality intervals, we compute ratios of cumulants using models of initial conditions, that can in this way be tested directly against experimental data on $v_n\{4\}/v_n\{2\}$.
To make our analysis as inclusive as possible, we test a wide variety of initial condition models, thus covering the spectrum of models typically used in hydrodynamic calculations.
Eventually, we employ these initial state parametrizations to predict $v_3\{6\}/v_3\{4\}$ in Pb+Pb collisions.
A similar study is carried over to high-multiplicity p+Pb collisions, in Sec.~\ref{sec:4}. 
Specifically, we employ the state-of-the-art Monte Carlo model of initial conditions for p+Pb collisions to make predictions for $v_3\{4\}/v_3\{2\}$, and $v_3\{6\}/v_3\{4\}$.

\section{cumulants and relative fluctuations}
\label{sec:2}
Anisotropic flow is the observation of a full spectrum of nonzero Fourier coefficients characterizing the azimuthal distribution of final-state particles in heavy-ion collisions.
Denoting the final-state azimuthal distribution by $P(\phi)$, its Fourier decomposition reads
\begin{equation}
\label{eq:vn}
P(\phi)=\frac{1}{2\pi}\sum_{n=-\infty}^{+\infty}V_n e^{-in\phi},
\end{equation}
and the quantity $v_n \equiv |V_n|$ is the coefficient of anisotropic flow in the $n$th harmonic.
In experiments, the number of final-state particles is not large enough to allow the computation of the Fourier series of Eq.~(\ref{eq:vn}) in every event. 
Flow coefficients are computed from azimuthal multi-particle correlations, which are averaged over many events. 
Since $P(\phi)$ is different in each collision, anisotropic flow coefficients fluctuate on an event-by-event basis.
Detailed information about the probability distribution of $v_n$ can be obtained by measuring its cumulants. 
A cumulant of order $m$ involves $m$-particle correlations, as well as lower order correlations~\cite{Borghini:2001vi,Bilandzic:2010jr,DiFrancesco:2016srj}: 
It is constructed by an order-by-order subtraction of trivial contributions coming from lower-order correlations. 
Cumulants are considered the best signature of the collective origin of anisotropic flow in heavy-ion collisions.
Nonzero values of higher-order cumulants have been measured in a wide range of collision systems, from Pb+Pb to p+p collisions \cite{Aad:2014vba,Khachatryan:2015waa,Khachatryan:2016txc}.

The cumulants of the distribution of $v_n$ are combinations of moments. 
Explicit expressions up to order 8 are~\cite{Giacalone:2016eyu}
\begin{align}
\nonumber v_n\{2\}^2 &= \bra v_n^2 \ket, \\
\nonumber v_n\{4\}^4 &= 2 \bra v_n^2\ket^2 - \bra v_n^4\ket, \\
\nonumber v_n\{6\}^6 &= \frac{1}{4} \biggl [ \bra v_n^6 \ket - 9 \bra v_n^2\ket \bra v_n^4 \ket + 12 \bra v_n^2\ket^3 \biggr], \\
\nonumber v_n\{8\}^8 &= \frac{1}{33} \biggl [ 144 \bra v_n^2\ket^4 - 144 \bra v_n^2\ket^2 \bra v_n^4 \ket + 18 \bra v_n^4\ket^2 \\
\label{eq:cum} &~~~~~~~~~~~~~~~~+ 16 \bra v_n^2 \ket \bra v_n^6 \ket - \bra v_n^8 \ket \biggr ],
\end{align}
where angular brackets denote an average over collision events in a given centrality class.
Cumulants are defined in such a way that $v_n\{2k\}=v_n$, if $v_n$ is the same for all events. 

Any quantity which is linearly proportional to $v_n$ has the same cumulants as $v_n$, up to a global factor.
If the scaling between $v_n$ and $\varepsilon_n$ were exactly linear, then, for any even integers $\mu$ and $\nu$ \cite{Ma:2016hkg},
\begin{equation}
\label{eq:ratio}
\frac{v_n\{\mu\}}{v_n\{\nu\}}=\frac{\varepsilon_n\{\mu\}}{\varepsilon_n\{\nu\}}.
\end{equation}
Ratios of cumulants quantify the relative fluctuations of $v_n$, which are equal to the relative fluctuations of $\varepsilon_n$ if the scaling is linear~\cite{Bhalerao:2011yg,Renk:2014jja}. 
In this work, we mainly focus on the ratio $v_n\{4\}/v_n\{2\}$ as a measure of the relative fluctuations of $v_n$. 
This ratio depends on the event-by-event fluctuations of $v_n$.
In particular, the larger the fluctuations of $v_n$ are, the smaller the ratio $v_n\{4\}/v_n\{2\}$ is. 
Higher-order ratios of cumulants, such as $v_n\{6\}/v_n\{4\}$, probe the non-Gaussianity of the fluctuations~\cite{Voloshin:2007pc,Giacalone:2016eyu}. 

Ratios of cumulants are interesting because they are independent of the hydrodynamic response (the proportionality coefficient between $\varepsilon_n$ and $v_n$), which is an important source of uncertainty when trying to constrain models of initial conditions from experimental data~\cite{Retinskaya:2013gca}.
Equation~(\ref{eq:ratio}) allows us to directly relate experimental data (left-hand side) to models of initial conditions (right-hand side).\footnote{A similar analysis was recently carried out at Relativistic Heavy Ion Collider (RHIC) energies within the AMPT model \cite{Ma:2016hkg}.}
The approximate linearity of the relation between $v_n$ and $\varepsilon_n$ in event-by-event hydrodynamics is typically measured using scatter plots~\cite{Niemi:2012aj} or the Pearson correlation coefficient~\cite{Gardim:2011xv}. 
Nevertheless, these approaches do not give any information on ratios of cumulants, and on the accuracy of Eq.~(\ref{eq:ratio}). 
More precisely, if one models the deviation from linear scaling by a Gaussian noise, $v_n=\kappa_n\varepsilon_n+\delta$, where $\delta$ is a random fluctuation with a Gaussian distribution, this noise will typically contribute to the rms value of $v_n\{2\}$, not to higher-order cumulants. 
Therefore, it is not at all trivial that ratios of cumulants are preserved by the hydrodynamic evolution. 
In the next section, we analyze the validity of Eq.~(\ref{eq:ratio}) more robustly, by testing this equation directly through hydrodynamic calculations. 

\section{$\textup{Pb+Pb}$ collisions}
\label{sec:3}

\begin{figure}[t!]
\centering
\includegraphics[width = \linewidth]{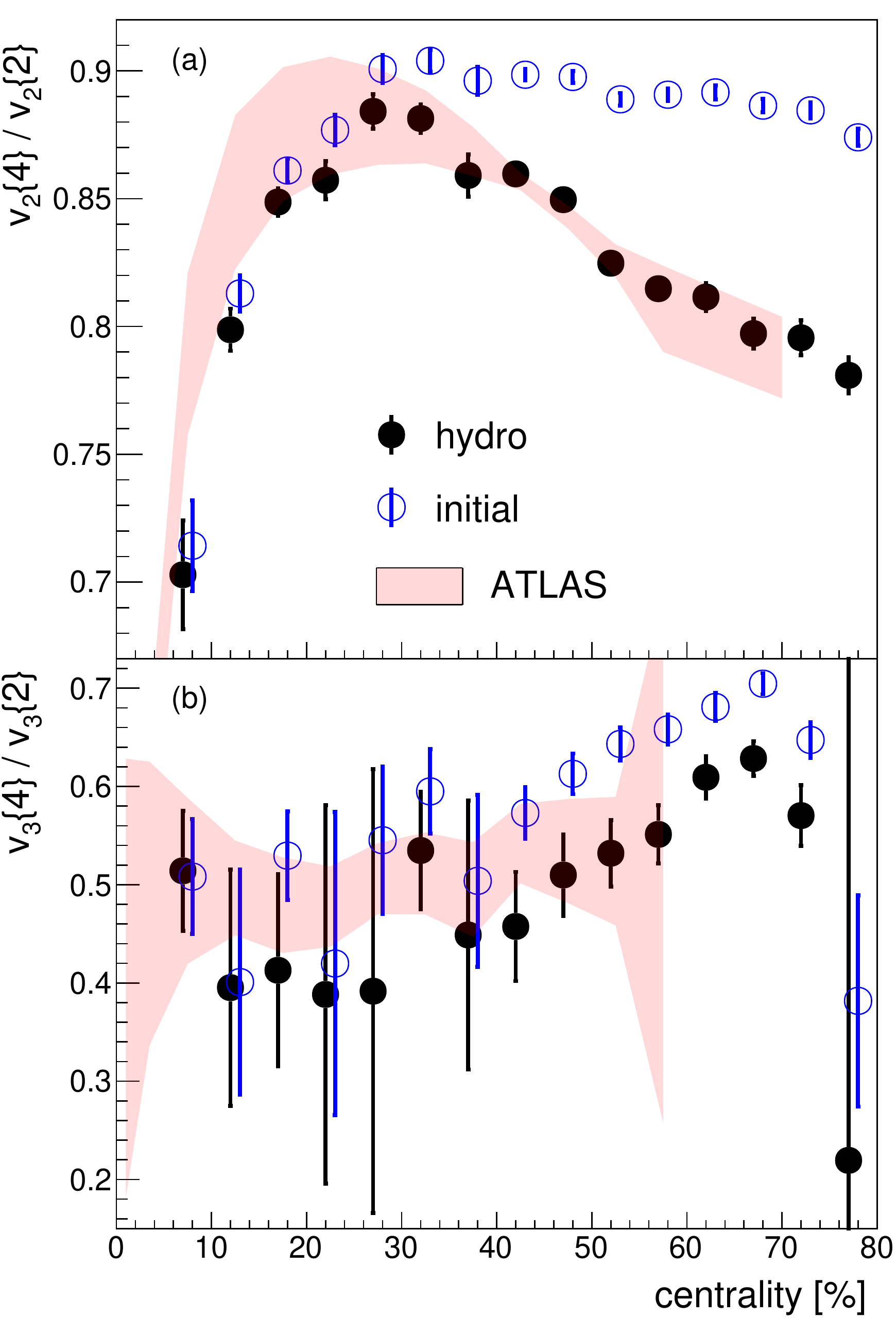}
\caption{(color online) Comparison between $v_n\{4\}/v_n\{2\}$  computed in hydrodynamics (full symbols) and $\varepsilon_n\{4\}/\varepsilon_n\{2\}$  computed from the corresponding initial energy density profiles (open symbols), for 2.76~TeV Pb+Pb collisions. Shaded bands: ATLAS data for $v_n\{4\}/v_n\{2\}$~\cite{Aad:2014vba}. Symbols are shifted horizontally for readability. (a) Elliptic flow ($n=2$). (b) Triangular flow ($n=3$).}
\label{fig:1}
\end{figure}
We first test the validity of Eq.~(\ref{eq:ratio}) for $v_2\{4\}/v_2\{2\}$ and $v_3\{4\}/v_3\{2\}$, by computing both sides of the equation in event-by-event hydrodynamics. 
We run hydrodynamic simulations of Pb+Pb collisions at ~$\sqrt[]{s}=2.76$~TeV.
The initial conditions from which initial anisotropies are computed are given by a Glauber Monte Carlo model~\cite{Loizides:2014vua,Rybczynski:2013yba}.
Initial density profiles are evolved by means of the viscous relativistic hydrodynamical code {\footnotesize V-USPHYDRO} \cite{Noronha-Hostler:2013gga,Noronha-Hostler:2014dqa,Noronha-Hostler:2015coa}.
We implement a shear viscosity over entropy ratio of $\eta/s=0.08$~\cite{Policastro:2001yc}, and we compute flow coefficients at freeze-out~\cite{Teaney:2003kp} for pions in the transverse momentum range $0.2<p_t<3$~GeV/$c$.
We compute $v_2\{4\}/v_2\{2\}$ and $v_3\{4\}/v_3\{2\}$ as function of centrality percentile.
Between 1000 and 5000 events are simulated in each centrality window, each event corresponding to a different initial geometry. 
Results are shown in Fig.~\ref{fig:1}, and are compared to the measurements of the ATLAS Collaboration~\cite{Aad:2014vba}. 
A first remark is that $v_3\{4\}/v_3\{2\}$ is smaller than $v_2\{4\}/v_2\{2\}$.
This means that $v_3$ fluctuations are larger than $v_2$ fluctuations, as expected since $v_3$ is solely due to fluctuations~\cite{Alver:2010gr}.
The smallness of $v_3\{4\}$ explains the large statistical error on the corresponding ratio. 
We now discuss, in turn, $v_2\{4\}/v_2\{2\}$ and $v_3\{4\}/v_3\{2\}$. 
In the centrality intervals where Eq.~(\ref{eq:ratio}) holds to a good approximation, we test initial condition models against experimental data.

\subsection{Elliptic flow fluctuations}
We start with $v_2$ [Fig.~\ref{fig:1}--(a)].
Equation~(\ref{eq:ratio}) holds approximately up to $20-30\%$ centrality, and gradually breaks down as the centrality percentile increases. 
The difference between $\varepsilon_2\{4\}/\varepsilon_2\{2\}$ and $v_2\{4\}/v_2\{2\}$ can be attributed to a cubic response term, proportional to $(\varepsilon_2)^3$~\cite{Noronha-Hostler:2015dbi}.
Once this nonlinear hydrodynamic response is taken into account, agreement with ATLAS data is excellent all the way up to 70\% centrality. 
As we shall explain below, a similar nonlinear hydrodynamic response is also needed for other models of initial conditions in order to match experimental data. 

\begin{figure}[t!]
\centering
\includegraphics[width =\linewidth]{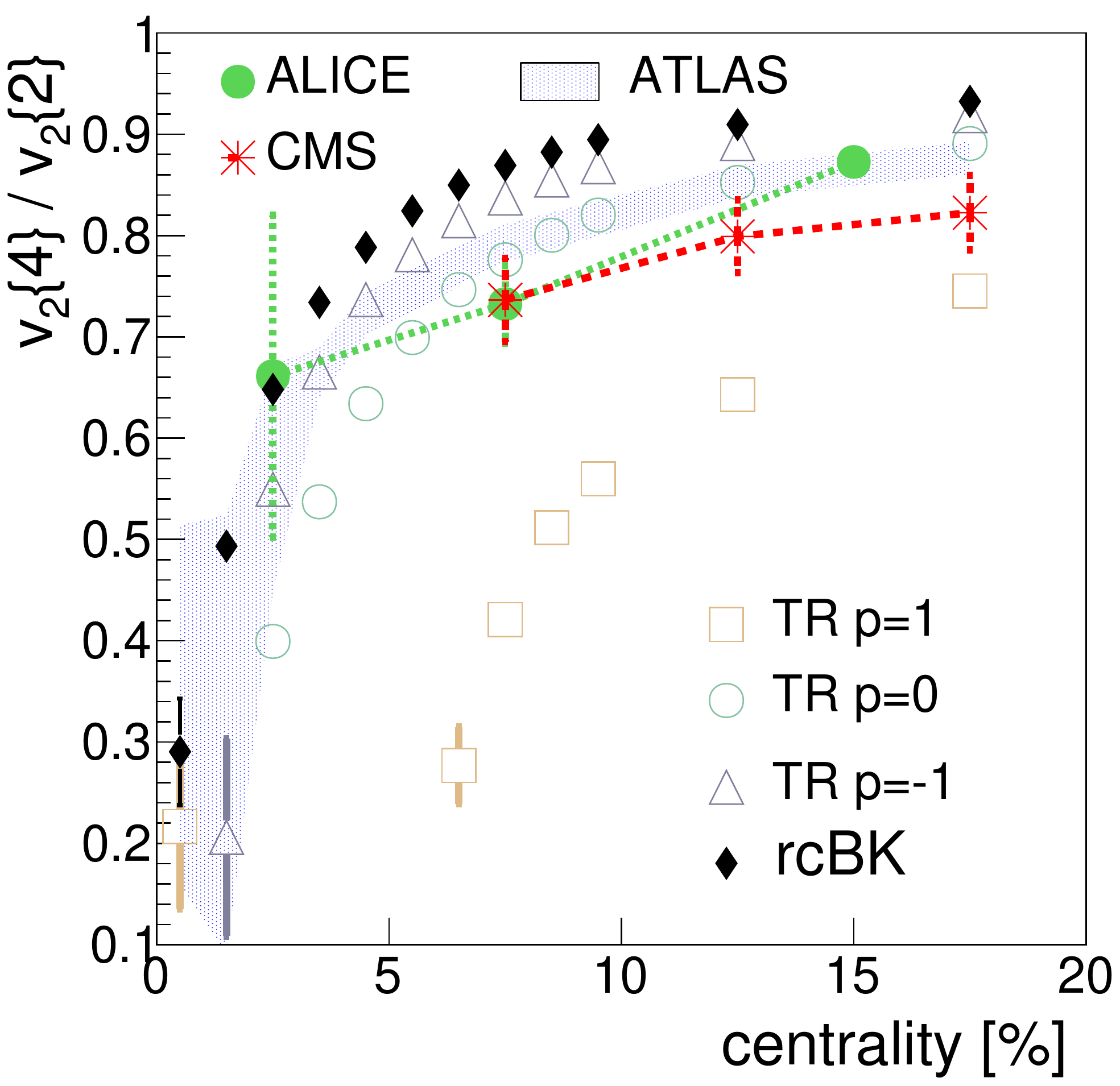}
\caption{(color online) Test of initial condition models using $v_2\{4\}/v_2\{2\}$ measured in Pb+Pb collisions at $2.76$~TeV up to 20\% centrality.  
Stars: CMS data \cite{Chatrchyan:2012ta}. Full circles: ALICE data \cite{ALICE:2011ab}. Shaded band: ATLAS data  \cite{Aad:2013xma}. 
Open symbols: Values of $\varepsilon_2\{4\}/\varepsilon_2\{2\}$ given by the \trento{} model with $p=-1$ (triangles), $p=0$ (circles) and $p=1$ (squares). Full diamonds: $\varepsilon_2\{4\}/\varepsilon_2\{2\}$ from the Monte Carlo rcBK model.}
\label{fig:2}
\end{figure}
Between 0\% and 20\% centrality, Eq.~(\ref{eq:ratio}) holds to a good approximation. 
Therefore, in this centrality window, the ratio $\varepsilon_2\{4\}/\varepsilon_2\{2\}$ provided by initial condition models can be tested directly against experimental data for $v_2\{4\}/v_2\{2\}$. 
We test the sensitivity of this observable to initial conditions using \trento{}~\cite{Moreland:2014oya}, a flexible parametric Monte Carlo model which effectively encompasses most of existing initial condition models~\cite{Bernhard:2016tnd}. 
The initial entropy density in \trento{} is expressed in terms of thickness functions, $T_A$ and $T_B$, associated with each of the colliding nuclei. 
Each thickness function is a sum of Gaussians, centered around the participant nucleons. 
The weight of each participant nucleon is a random variable, so that the contribution of a participant to the deposited energy density may fluctuate. 
The strength of these fluctuations is regulated by a parameter, $k$ (see the Appendix for details).
Another parameter is the width of the Gaussians, $\sigma$. 
The initial density profile is assumed to be a homogeneous function of degree 1 of the thickness functions $T_A$ and $T_B$, and a third parameter $p$ specifies this dependence. 
The values $p=1$, $p=0$, and $p=-1$ correspond respectively to an arithmetic mean, $(T_A+T_B)/2$, a geometric mean, $\sqrt{T_AT_B}$, and a harmonic mean, $T_AT_B/(T_A+T_B)$. 
The case $p=1$ corresponds to the Glauber Monte Carlo model, where the energy density is proportional to the number of wounded nucleons~\cite{Miller:2007ri}. 
The case $p=0$ gives results close to QCD-inspired models such as IP-Glasma~\cite{Schenke:2012wb,Moreland:2014oya} and EKRT~\cite{Niemi:2015qia,Bernhard:2016tnd}, while $p=-1$ is closer to the MC-KLN model~\cite{Drescher:2006ca,Bernhard:2016tnd}. 

We have checked that both $\varepsilon_2\{4\}/\varepsilon_2\{2\}$ and $\varepsilon_3\{4\}/\varepsilon_3\{2\}$, in Pb+Pb collisions, depend little on the parameters $k$ and $\sigma$. 
Therefore, we fix these parameters to the values suggested by the authors of \trento{} \cite{Moreland:2014oya}, which allow for a good description of the multiplicity distributions~\cite{Moreland:2014oya,Zakharov:2016gyu}. 
On the other hand, ratios of cumulants strongly depend on the third parameter, $p$. 
Results for $\varepsilon_2\{4\}/\varepsilon_2\{2\}$ are shown in Fig.~\ref{fig:2}, where they are compared to available experimental data on $v_2\{4\}/v_2\{2\}$.
The case $p=1$, corresponding to wounded nucleon scaling, is in poor agreement with data. 
In particular, the ratio $\varepsilon_2\{4\}/\varepsilon_2\{2\}$ is below data. 
This means that the relative fluctuations of $\varepsilon_2$ are too large, causing $\varepsilon_2\{4\}$ to fall too steeply in central collisions~\cite{Bhalerao:2011yg}.
The other values of $p$, $p=0$, and $p=-1$, corresponding to saturation models, are in fair agreement with data.\footnote{A comparison of the behaviors of $v_2\{2\}$ and $\varepsilon_2\{2\}$ in the $0-5\%$ centrality range also shows that the MC-KLN model is in better agreement with data than the Glauber model~\cite{ALICE:2011ab}.} 
Note that, in central collisions, $\varepsilon_2\{4\}$ is essentially equal to the mean eccentricity in the reaction plane~\cite{Giacalone:2016eyu}.
Saturation-inspired models are known to predict a larger mean eccentricity in the reaction plane than the Glauber model~\cite{Hirano:2005xf,Lappi:2006xc}. 
The larger mean eccentricity implies that relative fluctuations of $\varepsilon_2$ are smaller. 
Therefore, the ratio $\varepsilon_2\{4\}/\varepsilon_2\{2\}$ is larger. 

Figure~\ref{fig:2} also displays, for comparison, results obtained using the Monte Carlo rcBK~\cite{Albacete:2010ad} initial state model.
This QCD-inspired model predicts a mean eccentricity in the reaction plane comparable to the MC-KLN model~\cite{Retinskaya:2013gca}, which explains why results are similar to \trento{} with $p=-1$. 

Above 20\% centrality (not shown in figure), we find that all models overpredict $v_2\{4\}/v_2\{2\}$, much as in Fig.~\ref{fig:1} (a).
Therefore, for mid-central and peripheral collisions, all parameterizations of initial conditions require a nonlinear hydrodynamic response, breaking Eq.~(\ref{eq:ratio}), in order to be compatible with data.\footnote{A similar conclusion was drawn from simulations within the IP Glasma model~\cite{Schenke:2013aza}.}

\begin{figure*}[t!]
\centering
\includegraphics[width=.9\linewidth]{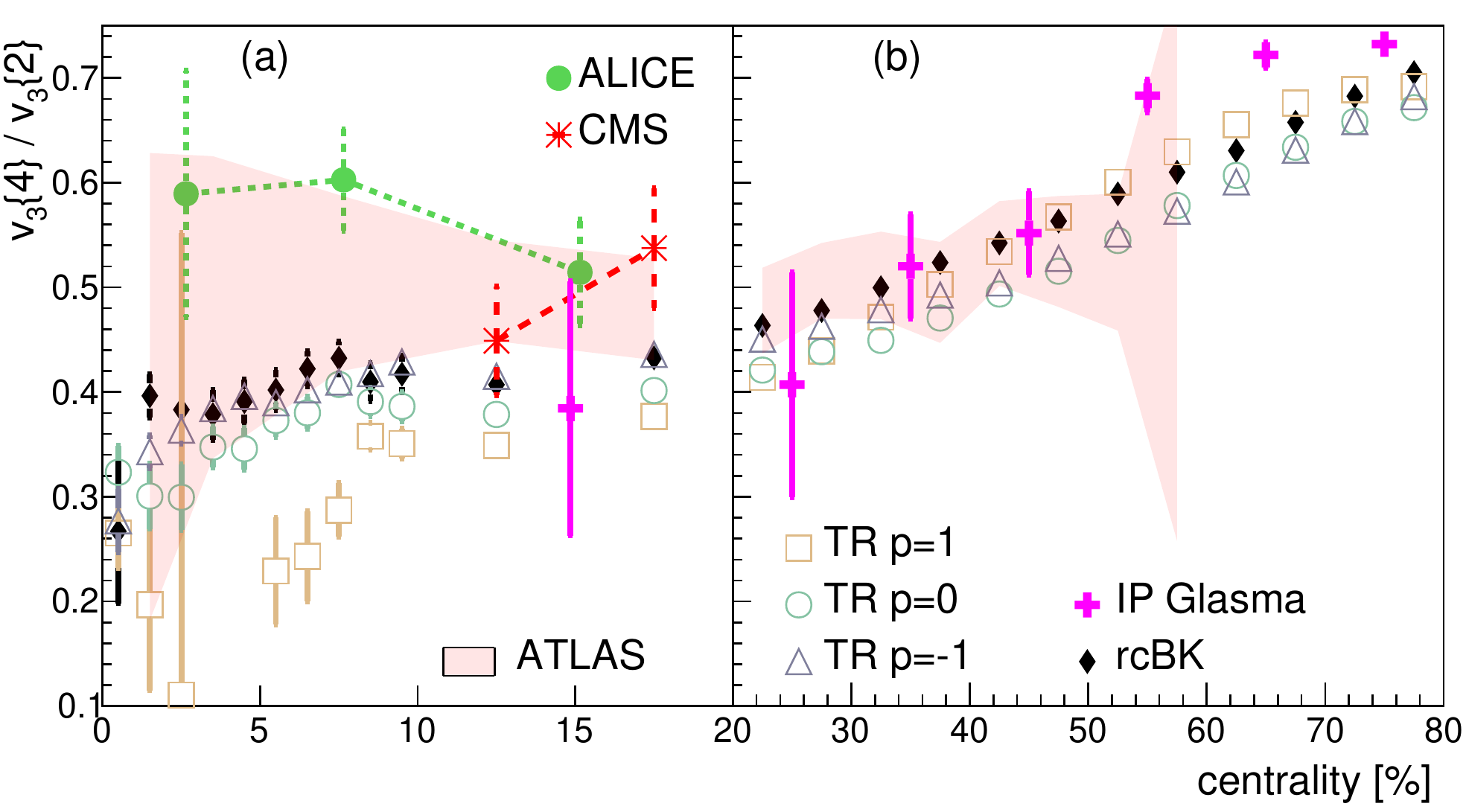}
\caption{(color online) Test of initial condition models using $v_3\{4\}/v_3\{2\}$ measured in 2.76~TeV Pb+Pb collisions: (a) up to 20\% centrality; (b) between 20\% and 80\% centrality. Stars: CMS data \cite{Chatrchyan:2013kba}. Full circles: ALICE data \cite{ALICE:2011ab}. Shaded band: ATLAS data \cite{Aad:2014vba}. 
ALICE and CMS data are not shown in panel (b) for the sake of readability, but are compatible with ATLAS data.  Remaining symbols correspond to values of $\varepsilon_3\{4\}/\varepsilon_3\{2\}$ from several models of initial conditions. Open symbols: \trento{}, with $p=-1$ (triangles), $p=0$ (circles), $p=1$ (squares). Full crosses: IP-Glasma~\cite{Schenke:2012wb}. Full diamonds: Monte Carlo rcBK~\cite{Albacete:2010ad}.}
\label{fig:3}
\end{figure*}

\subsection{Triangular flow fluctuations}
We now test the validity of Eq.~(\ref{eq:ratio}) in the case of triangular flow fluctuations. 
Hydrodynamic results in Fig.~\ref{fig:1}~(b) show that, as in the case of elliptic flow,  $\varepsilon_3\{4\}/\varepsilon_3\{2\}$ is systematically larger than $v_3\{4\}/v_3\{2\}$ above 40\% centrality.
This can again be attributed to a nonlinear hydrodynamic response, whose effect is, however, smaller for $v_3$ than for $v_2$. 
A possible explanation to this nonlinear effect could be a coupling between $v_2$ and $v_1$~\cite{Gardim:2014tya}. 
In general, one expects any nonlinear effect to be associated with the large magnitude of $v_2$, which is by far the largest Fourier harmonic~\cite{Teaney:2012ke}. 
Therefore, even though the large error bars in Fig.~\ref{fig:1}~(b) prevent any definite conclusion, we expect the nonlinear response between $\varepsilon_3$ and $v_3$ to be small in central collisions. 

By virtue of this conclusion, we compare $v_3\{4\}/v_3\{2\}$ from experimental data to $\varepsilon_3\{4\}/\varepsilon_3\{2\}$ from initial state models, across the full centrality range.
We implement the same models as in Fig.~\ref{fig:2}, and we also show results obtained using the IP-Glasma~\cite{Schenke:2012wb} model, for sake of comparison. 
Results are displayed in Fig.~\ref{fig:3}, where the $0-20\%$ centrality range is zoomed in [panel (a)] for readability.
A first remark is that experimental data do not exhibit any clear dependence on centrality. 
Relative $\varepsilon_3$ fluctuations, on the other hand, grow from central to peripheral collisions in all the tested models.
This centrality dependence has a simple explanation: Since the system size decreases as a function of the centrality percentile, the relative fluctuations of $\varepsilon_3$ become larger~\cite{Bhalerao:2011bp}.  
In general, the nonlinear hydrodynamic response seen in  Fig.~\ref{fig:1}--(b) would help in decreasing $v_3\{4\}/v_3\{2\}$  above 40\% centrality and reducing the centrality dependence, which is seen in models and not in data. 
However, all configurations of \trento{} in Fig.~\ref{fig:3}--(b) are compatible with ATLAS data above 40\% centrality, and some points would fall below data if a nonlinear response were included. 

Figure~\ref{fig:3}--(a) presents results in the 20\% most central collisions, where we use a finer centrality binning for initial-state models. 
In this centrality range, we do not foresee any significant nonlinear hydrodynamic response, and initial state calculations should match data. 
Data points (in particular the measurements of the ALICE Collaboration) are, however, above the predictions of all models. 
As observed for elliptic flow, the wounded nucleon prescription (p=1) gives the worst results. 
We conclude that initial state models overestimate the relative fluctuations of $\varepsilon_3$ in central Pb+Pb collisions.
 
\subsection{Predictions for $\mathbf{v_3\{6\}}$}
\begin{figure}[t!]
\centering
\includegraphics[width = \linewidth]{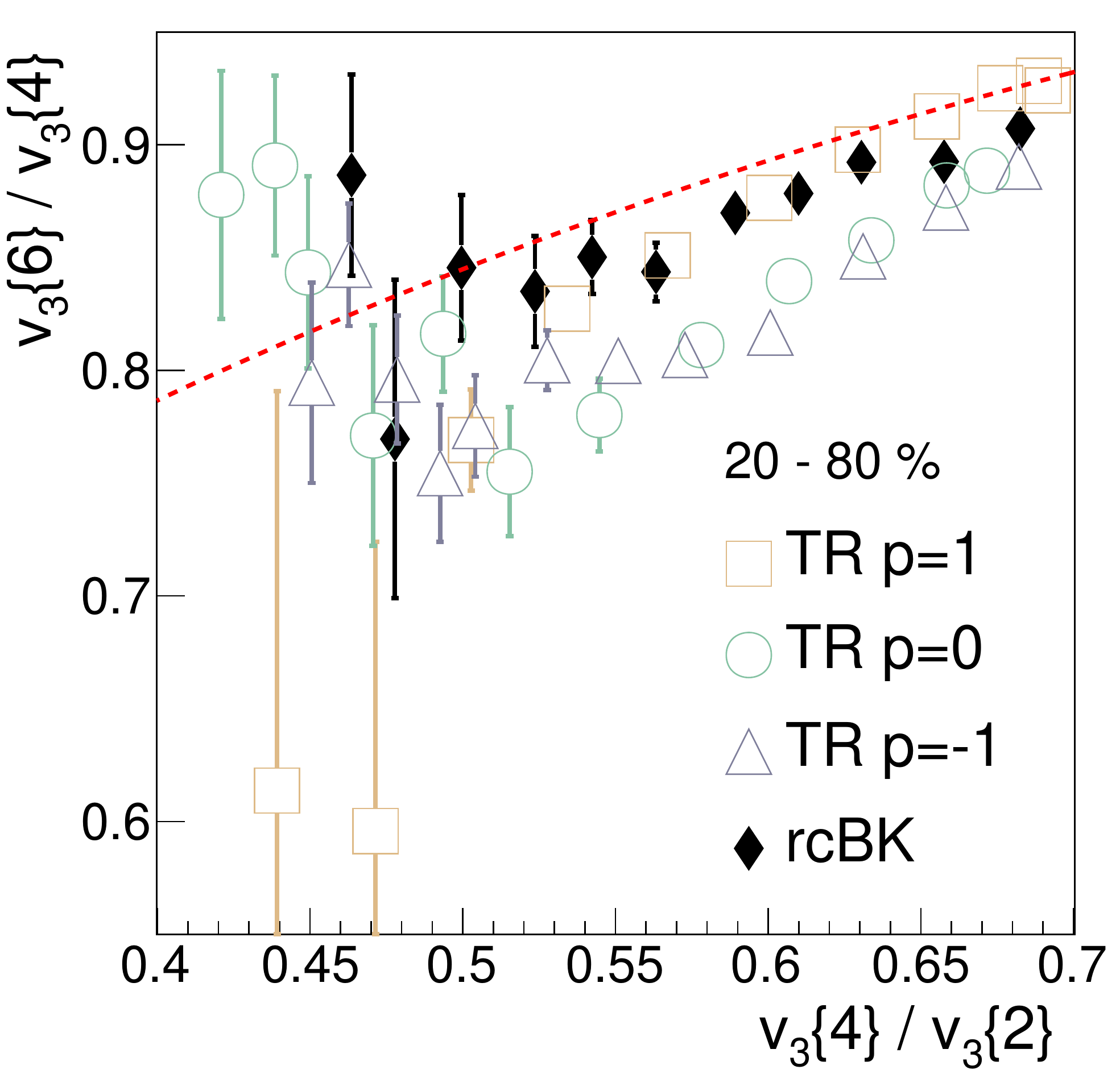}
\caption{(color online) Predictions for $v_3\{6\}/v_3\{4\}$ in 2.76~TeV Pb+Pb collisions, from several models of initial conditions, in the $20-80$\% centrality range. Empty symbols: Predictions of \trento{} with $p=1$ (squares), $p=0$ (circles), and $p=-1$ (triangles). Full diamonds: Prediction of Monte Carlo rcBK \cite{Albacete:2010ad}. The red dashed line is the prediction of the power distribution \cite{Yan:2013laa}.}
\label{fig:4}
\end{figure}
We now use Eq.~(\ref{eq:ratio})  to make predictions for  $v_3\{6\}$ in Pb+Pb collisions.  
The number of events in our hydrodynamic calculations is not large enough to test directly the validity of Eq.~(\ref{eq:ratio}) for $v_3\{6\}/v_3\{4\}$. 
However, we have noted that the nonlinear hydrodynamic response is smaller for $v_3$ than for $v_2$. 
In addition, a previous study~\cite{Giacalone:2016eyu} has shown that, even for $v_2$, the ratio $v_2\{6\}/v_2\{4\}$ is little affected by the nonlinear response, so that  Eq.~(\ref{eq:ratio})  applies, to a good approximation, up to very peripheral collisions. 
Therefore, we assume that Eq.~(\ref{eq:ratio}) yields a reasonable estimate of $v_3\{6\}/v_3\{4\}$, and we make predictions on this basis using our \trento{} configurations and the rcBK model. 

It has been argued  that the probability distribution of $\varepsilon_3$~\cite{Bravina:2015sda}, which is solely due to fluctuations, is well described by the power distribution~\cite{Yan:2013laa}, which has a single  free parameter characterizing the rms value of $\varepsilon_3$.
If the distribution of $\varepsilon_3$ follows the power distribution, then, the ratio $\varepsilon_3\{6\}/\varepsilon_3\{4\}$ is a simple function of  the ratio $\varepsilon_3\{4\}/\varepsilon_3\{2\}$, which is displayed as a dashed line in Fig.~\ref{fig:4}. 
By running Monte Carlo simulations of the initial state, we can test whether the results fall on this line. 
To this purpose, we simulate a large number of initial conditions for Pb+Pb collisions, and we compute $\varepsilon_3\{6\}/\varepsilon_3\{4\}$ in the $20-80\%$ centrality range. 

Results are shown as symbols in Fig.~\ref{fig:4}. 
The centrality percentile corresponding to each symbol can be inferred from Fig.~\ref{fig:3} (b). 
For a given model, $\varepsilon_3\{4\}/\varepsilon_3\{2\}$ increases with the centrality percentile.  
The rcBK model agrees with the prediction of the power distribution, while the various parametrizations of the Trento model give in general values of $\varepsilon_3\{6\}/\varepsilon_3\{4\}$ which fall below the expected curve. 
The fact that the power distribution can be a poor approximation for large systems such as Pb+Pb collisions, even if the anisotropy is solely due to fluctuations, has already been pointed out in Ref.~\cite{Gronqvist:2016hym}. 
Even though precise figures depend on the particular model used, we predict on the basis of our Monte Carlo calculations, and of Eq.~(\ref{eq:ratio}), that $v_3\{6\}/v_3\{4\}$ should lie between 0.75 and 0.85 in the $30-50\%$ centrality range. 

\section{High-multiplicity $\textup{p+Pb}$ collisions}
\label{sec:4}

In this Section, we study relative flow fluctuations in high-multiplicity p+Pb collisions at $\sqrt{s} = 5.02 ~\textup{TeV}$, and we make quantitative predictions for higher-order cumulants of $v_2$ and $v_3$. 
Nonzero elliptic and triangular flow values have been measured in p+Pb systems~\cite{Aad:2013fja,Chatrchyan:2013nka,Abelev:2014mda,Khachatryan:2015waa}. 
In particular, a positive $v_2\{4\}^4$ has been reported by all collaborations, suggesting that the measured azimuthal correlations originate from a collective effect. 
Hydrodynamic simulations have also been carried out~\cite{Bozek:2013uha,Bzdak:2013zma,Qin:2013bha,Schenke:2014zha,Kozlov:2014fqa,Shen:2016zpp}, using either IP-Glasma or Glauber Monte Carlo initial conditions. 
Satisfactory agreement with data was found, which supports the hydrodynamic picture as a valid description of the p+Pb system~\cite{Weller:2017tsr}. 
Since elliptic flow is significantly smaller in p+Pb collisions than in Pb+Pb collisions~\cite{Abelev:2012ola}, one does not expect a significant nonlinear hydrodynamic response, and we assume that Eq.~(\ref{eq:ratio}) always holds. 
Event-by-event hydrodynamic simulations confirm that $v_2$ and $v_3$ scale linearly with the corresponding initial anisotropies, $\varepsilon_2$ and $\varepsilon_3$~\cite{Bozek:2013uha}.

We first select a model of initial conditions by requiring that it reproduces the first nontrivial ratio of cumulants, $v_2\{4\}/v_2\{2\}$, which has been measured by the CMS Collaboration \cite{Chatrchyan:2013nka}, as a function of centrality percentile.
As in the previous section, we employ the \trento{} model. 
However, the sets of parameters that give a reasonable description of Pb+Pb data fail to describe p+Pb data. 
Specifically, the values $p=-1$ and $p=0$, which provide a good description of experimental data in Fig.~\ref{fig:2}, yield a negative $\varepsilon_2\{4\}^4$ in p+Pb collisions (i.e., an undefined $\varepsilon_2\{4\}$), and values of $\varepsilon_2$ which are much smaller than needed in order to explain the magnitude of the measured $v_2$. 
This is due to the fact that, with these parameters, the initial density profile is always included in the transverse area spanned by the proton, which is circular. 
For the same reason, the IP-Glasma model underpredicts $v_2$ by a large factor, unless one allows the proton to be ``eccentric''~\cite{Schenke:2014zha}. 
On the other hand, previous hydrodynamic calculations have shown that the implementation of Glauber Monte Carlo initial conditions yields results in good agreement with p+Pb data. 
We therefore choose the value $p=1$, corresponding to the Glauber model, even though it does give a bad description of flow fluctuations in Pb+Pb data.  
We fix the parameter governing the multiplicity fluctuations to the value $k=0.9$~\cite{Bozek:2013uha}, and we have checked that the initial entropy distribution folded with a Poisson distribution yields the final multiplicity distribution observed in experiments~\cite{Moreland:2014oya}. 
We allow the width $\sigma$ of the source associated with each nucleon to vary. 
Previous calculations implement $\sigma=0.4$~fm. 
As we shall see, results depend somewhat on the value of $\sigma$. 

\begin{figure}[t!]
\centering
\includegraphics[width = \linewidth]{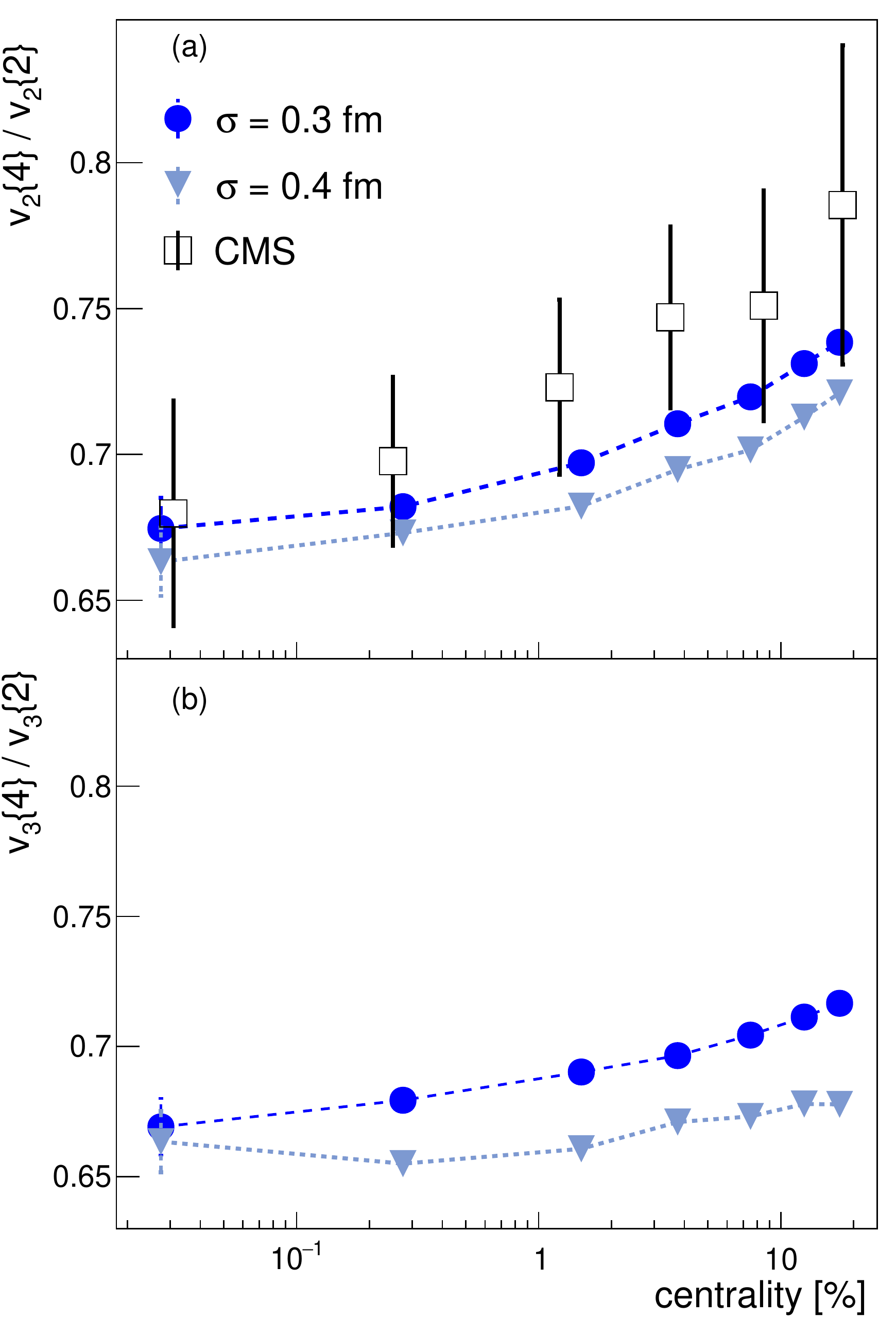}
\caption{(color online) $v_2\{4\} / v_2\{2\}$ (a) and $v_3\{4\} / v_3\{2\}$ (b) as functions of centrality percentile in 5.02~TeV p+Pb collisions. Full circles: \trento{} parametrization with $\sigma=0.3$~fm. Triangles: \trento{} parametrization with $\sigma=0.4$~fm.  Squares: CMS data \cite{Chatrchyan:2013nka}. The centrality binning of CMS data is taken from Table~I of Ref.~\cite{Kozlov:2014fqa}.}
\label{fig:5}
\end{figure}
Figure~\ref{fig:5}--(a) displays the comparison between $\varepsilon_2\{4\}/\varepsilon_2\{2\}$ from the \trento{} model, and $v_2\{4\}/v_2\{2\}$ measured by the CMS Collaboration~\cite{Chatrchyan:2013nka}.
The centrality percentile in our \trento{} configuration is defined from the multiplicity of produced particles, thus mimicking the experimental situation. 
For $\sigma=0.4$~fm, the model is compatible with experimental data in ultracentral collisions, but underestimates the ratio of cumulants as soon as the centrality percentile increases.
These results are consistent with the hydrodynamic results by Kozlov {\it et al.}  \cite{Kozlov:2014fqa}, who find a $v_2\{2\}$ which matches data, and a slightly underpredicted $v_2\{4\}$.
Agreement with experimental data mildly improves if the participant nucleons widths are lowered to $\sigma = 0.3$~fm.
Lower values of $\sigma$ yield more spiky initial density profiles, and are known to increase the magnitude of $\varepsilon_2$ and $\varepsilon_3$ in small systems~\cite{Zakharov:2016gyu}.
In central p+Pb collisions, we find that the rms $\varepsilon_2$ increases by  8\% when $\sigma$ is lowered from $0.4$~fm to $0.3$~fm (the rms $\varepsilon_3$ increases by 12\%). 
Larger values of $\varepsilon_n$ are known to yield larger values of $\varepsilon_n\{4\}/\varepsilon_n\{2\}$~\cite{Yan:2013laa}.
Even when $\sigma = 0.3$~fm, our parametrization of initial conditions tends to underpredict $v_2\{4\}/v_2\{2\}$.
Note, however, that the experimental measurements of $v_2\{4\}$ and $v_2\{2\}$ differ in the implementation, and the comparison with our results may not be consistent: 
$v_2\{2\}$ is measured with a large pseudorapidity ($\eta$) gap to suppress nonflow effects, but no $\eta$ gap is implemented in the measurement of $v_2\{4\}$.  
Therefore, measurements of $v_2\{4\}$ may be affected by nonflow, short-range (near side) correlations. 
In addition, the $\eta$ gap typically reduces $v_2\{2\}$, because of pseudorapidity dependent event-plane fluctuations~\cite{Khachatryan:2015oea}.
Recently, a novel method to measure multi particle cumulants in small systems was proposed \cite{Jia:2017hbm}.
It implements pseudorapidity gaps for the measurements of four-particle correlations.
The results reported by the authors of this method suggest that, in proton+proton collisions, the measured four-particle correlations ($v_2\{4\}$ and $v_3\{4\}$) may originate entirely from nonflow contributions.
We expect agreement between our model and experimental data to be improved if $v_2\{2\}$ and $v_2\{4\}$ are measured using the same sample of detected particles. 

We now make predictions for the ratio $v_3\{4\}/v_3\{2\}$, which has not yet been measured in p+Pb collisions.
$v_3\{4\}$ in p+Pb collisions has been computed in event-by-event hydrodynamics~\cite{Kozlov:2014fqa}. 
Nevertheless, the ratio $v_3\{4\} / v_3\{2\}$ is a more robust quantity, in the sense that depends little on model parameters (such as viscosity, or freeze-out temperature) and kinematic cuts ($p_t$)\footnote{The fact that the ratios of cumulants are not sensitive to the value of $\eta/s$ is clearly inferable from the results of \cite{Kozlov:2014fqa}. There, the authors show explicitly that both $v_2\{2\}$ and $v_2\{4\}$ increase (decrease) by the same amount when the value of $\eta/s$ is raised (lowered).}. 
Our results, from the \trento{} configuration with $p=1$, are shown in Fig.~\ref{fig:5}--(b).
We find $v_3\{4\} / v_3\{2\}$ to be slightly smaller than 
$v_2\{4\} / v_2\{2\}$ in Fig.~\ref{fig:5}--(a).
The sensitivity to the value of $\sigma$ is somewhat stronger for $v_3$ than for $v_2$.

\begin{figure}[!t]
\centering
\includegraphics[width = \linewidth]{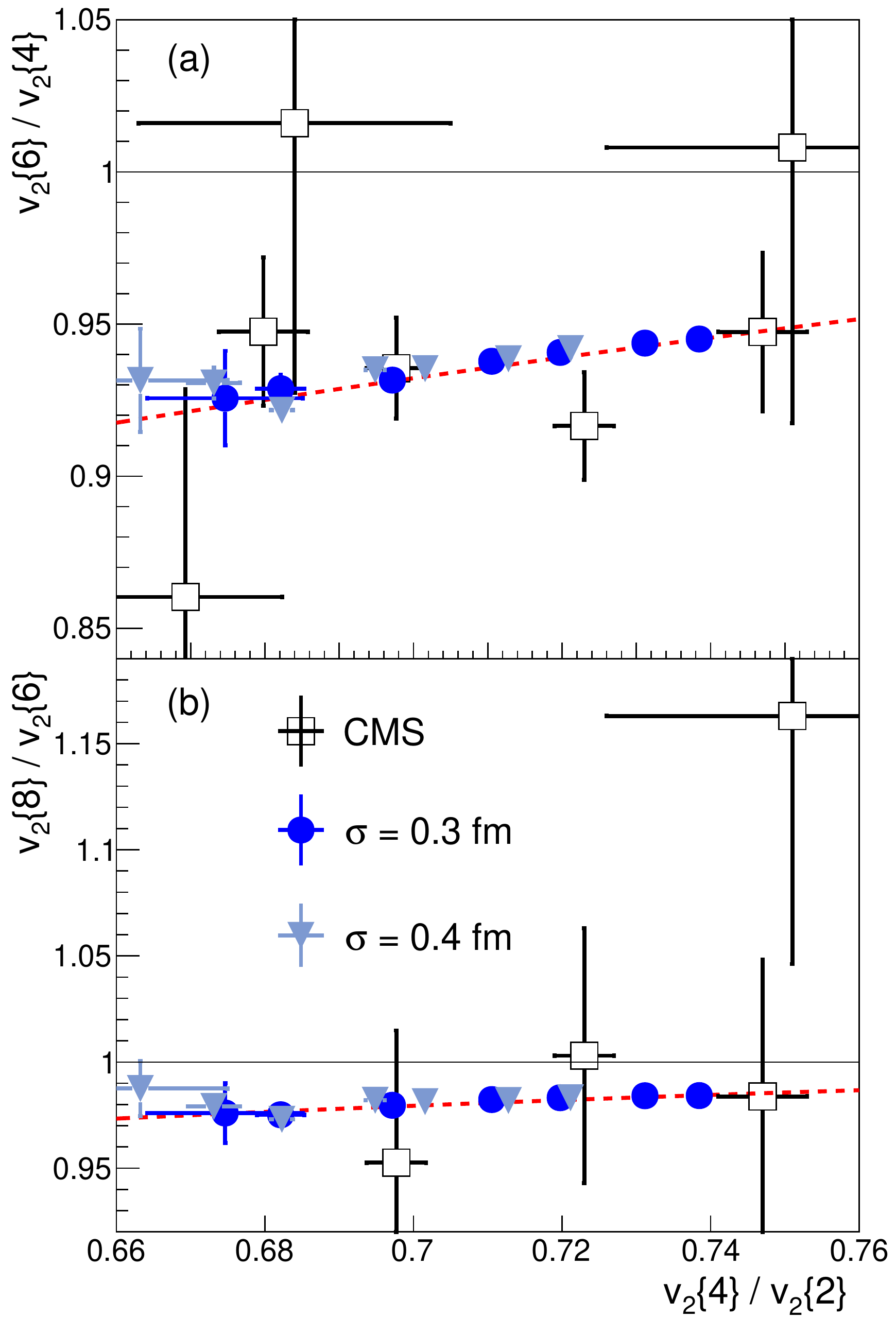}
\caption{(color online) Eccentricity-driven predictions for $v_2\{6\}/v_2\{4\}$ and $v_2\{8\}/v_2\{6\}$ as function of $v_2\{4\}/v_2\{2\}$ in 5.02~TeV p+Pb collisions. Full symbols: \trento{} parametrization with $\sigma=0.3$~fm (circles) and $\sigma=0.4$~fm (triangles). Empty symbols: CMS data \cite{Khachatryan:2015waa}. The red dashed line represents the prediction of the power distribution \cite{Yan:2013laa}.}
\label{fig:6}
\end{figure}

The CMS Collaboration has also measured $v_2\{6\}/v_2\{4\}$ and $v_2\{8\}/v_2\{6\}$~\cite{Khachatryan:2015waa} in p+Pb collisions. 
Our \trento{} results for these ratios are shown in Fig.~\ref{fig:6}.
As in Fig.~\ref{fig:4}, we plot them as a function of the lowest-order ratio, $v_2\{4\}/v_2\{2\}$.
We observe that our Monte Carlo results are in perfect agreement with the prediction of the power distribution (dashed line in Fig.~\ref{fig:6}). 
This confirms that the power distribution is a good description of eccentricity fluctuations in small systems, irrespective of the details of the simulated configurations~\cite{Gronqvist:2016hym}. 
Existing CMS data exhibit as well good agreement with this theoretical prediction. 
Future measurements with smaller error bars will provide a crucial test of the eccentricity-driven nature of $v_2$ in proton+nucleus collisions. 

\begin{figure}[t!]
\centering
\includegraphics[width = \linewidth]{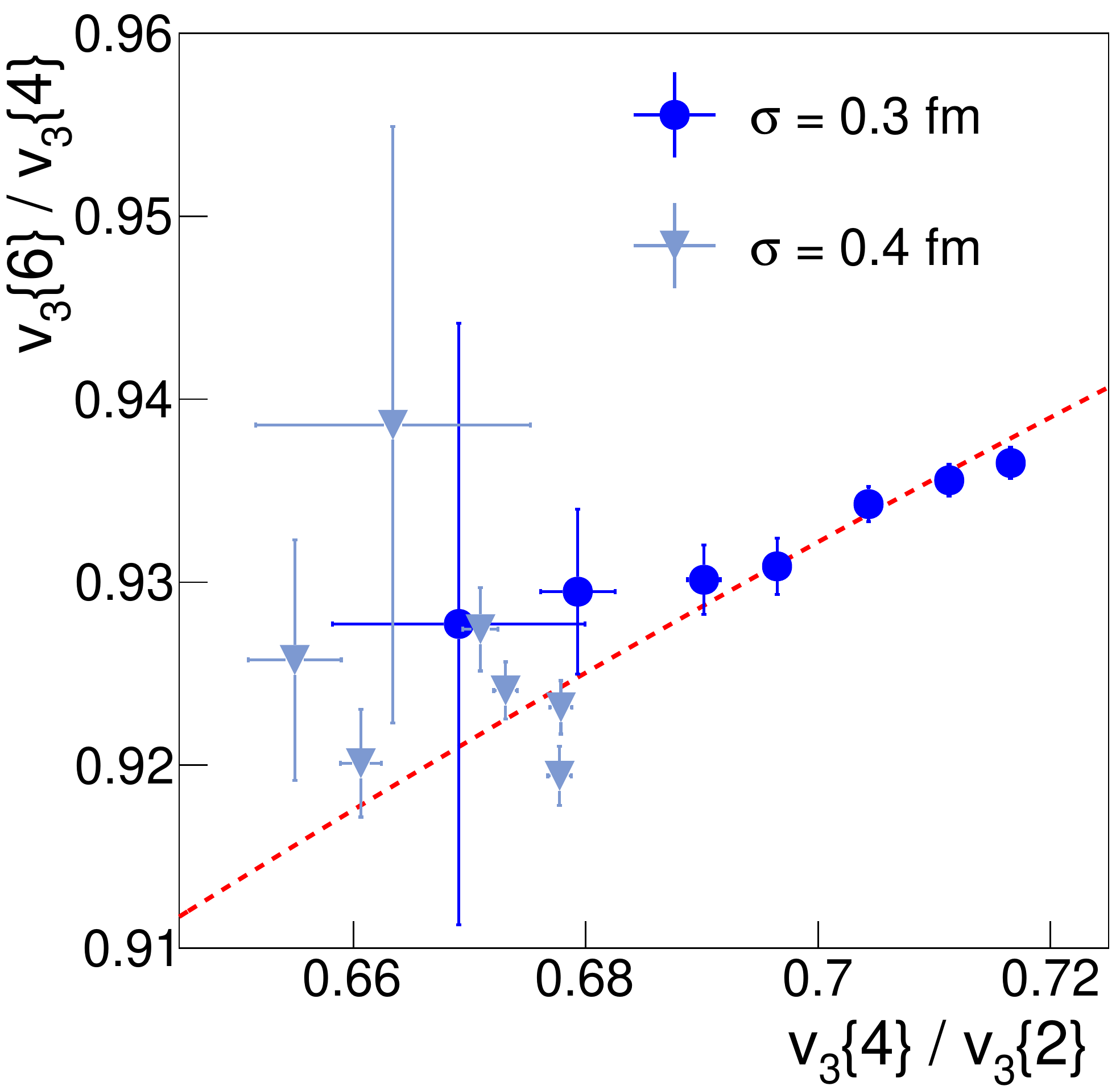}
\caption{(color online) Prediction for $v_3\{6\}/v_3\{4\}$ as function of $v_3\{4\}/v_3\{2\}$ in central 5.02~TeV p+Pb collisions, from different \trento{} parametrizations. Circles: $\sigma=0.3$~fm. Triangles: $\sigma=0.4$~fm. The red dashed line represents the prediction of the power distribution~\cite{Yan:2013laa}.}
\label{fig:7}
\end{figure}

Eventually, we make a prediction for $v_3\{6\}/v_3\{4\}$ as function of $v_3\{4\}/v_3\{2\}$ in central p+Pb collisions. 
Results are displayed in Fig.~\ref{fig:7}, for both $\sigma=0.3$~fm and $\sigma=0.4$~fm.
Our Monte Carlo results are well described by the power distribution, although with large error bars for $\sigma=0.4$~fm.  

\section{Discussion and outlook}
We have shown that ratios of cumulants are a powerful tool to test models of initial conditions directly against experimental data.
The Glauber Monte Carlo model, which is by far the most employed model in both experimental and theoretical analyses, is excluded by experimental data on elliptic flow fluctuations in central Pb+Pb collisions.
On the other hand, saturation models (mimicked by the \trento{} parametrizations with $p=0$ or $p=-1$) provide a good description of the experimental results.
However, even if these models predict the correct fluctuations of $v_2$, they overpredict the fluctuations of $v_3$ in central Pb+Pb collisions.
A possible explanation is that they overestimate both the fluctuations and the mean eccentricity, $\varepsilon_2$, in the reaction plane.
In this way, the error cancels in the ratio $v_2\{4\}/v_2\{2\}$, but not in the corresponding ratio for $v_3$, which is solely due to fluctuations. 
It will be of crucial importance to reduce the error bars on experimental data on $v_3\{4\}$ in central Pb+Pb collisions, in order to check whether the ratio $v_3\{4\}/v_3\{2\}$ is independent of centrality, as suggested by ALICE data.
Indeed, this observation does not seem compatible with existing models of initial conditions. 

The parametrizations of the initial state that are suitable for describing central Pb+Pb collisions, can not be employed in central p+Pb collisions, and vice versa. 
Indeed, the Glauber model, which is excluded by Pb+Pb data, provides the only reasonable description of p+Pb collisions. 
We do not consider this as a contradiction, because we are merely trying to identify the parametrization which captures the initial geometry in a given system, and we do not aim at a unified description of all systems. 
We predict that the ratio $v_3\{4\}/v_3\{2\}$ is very close to $v_2\{4\}/v_2\{2\}$ in high-multiplicity p+Pb collisions, and both the distributions of $v_2$ and $v_3$ to follow the power distribution.
These results imply that, up to small corrections, the same non-Gaussianities drive the fluctuations of $\varepsilon_2$ and $\varepsilon_3$.
Our explicit test of the power behavior up to higher-order cumulants, in particular, suggests that the main non-Gaussianity driving the fluctuations is the fact that the distributions are bounded by unity.
However, nonflow effects differ for $v_2$ and $v_3$ (back-to-back correlations typically increase $v_2$, and decrease $v_3$) and must be carefully removed in the analysis.

As a final remark, we stress that the conclusions drawn in our p+Pb analysis should hold in any small system model where $\varepsilon_2$ and $\varepsilon_3$ originate solely from fluctuations. 
It would be rather natural, then, to extend this analysis to the case of high-multiplicity proton+proton collisions, where the observed azimuthal multi particle correlations hint at the onset of collective effects~\cite{Aad:2015gqa,Khachatryan:2016txc}.
These new data have triggered novel models of initial conditions~\cite{Loizides:2016djv,Welsh:2016siu}, which can be tested against experimental data using ratios of cumulants, as done in this work for p+Pb collisions.

\section*{Acknowledgements}
J.N.H. acknowledges the use of the Maxwell Cluster and the advanced support from the Center of Advanced Computing and Data Systems at the University of Houston and was supported by the National
Science Foundation under Grant No. PHY-1513864.
We thank Matt Luzum for useful discussions.
G.G. wishes to thank Scott Moreland for kind assistance with the use of \trento{}.

\appendix
\section{The \trento{} model}
\label{s:trento}
\trento{} is a flexible parametric Monte Carlo model for the initial conditions of heavy-ion collisions, which encompasses several other models of initial conditions~\cite{Moreland:2014oya}.
Consider the case of a nucleon A colliding with a nucleon B.
Each participant nucleon deposits entropy in the transverse plane according to a Gaussian distribution of width $\sigma$, which reads
\begin{equation}
S_{A,B} = w_\textup{A,B} ~\frac{1}{2\pi\sigma^2} \exp \biggl [ \frac{(x-x_\textup{A,B})^2+(y-y_\textup{A,B})^2}{2\sigma^2} \biggr].
\end{equation}
The normalization, $w$, is a random number which is assigned to each participant nucleon.
Its probability distribution is a $\Gamma$ distribution, whose mean value is equal to unity, and whose width is regulated by a parameter, $k$.
The total initial entropy profile is computed through a generalized average of Gaussian sources,
\begin{equation}
\label{eq:p}
S(p;S_\textup{A},S_\textup{B}) = \biggl ( \frac{S_\textup{A}^p + S_\textup{B}^p}{2} \biggr )^{\frac{1}{p}},
\end{equation}
where $p$ is an arbitrary real parameter.
The previous formula can be generalized to the case of a nucleus A colliding with a nucleus B \cite{Moreland:2014oya}.
Note that, for $p=1$, nuclear density profiles are superimposed ($S \propto S_{\rm A}+S_{\rm B}$).
If $p=0$ or $p=-1$, instead, the initial entropy deposition is computed through the product of the two nuclear density profiles ($S \propto S_{\rm A}S_{\rm B}$).
Varying the value of $p$, it is possible to construct initial entropy profiles according to different prescriptions \cite{Bernhard:2016tnd}: $p=1$ is the wounded nucleon model; lower values of $p$ reproduce QCD-based models, such as EKRT \cite{Eskola:2001bf} ($p=0$), or Monte Carlo KLN \cite{Kharzeev:2004if} ($p=-0.67$).

\end{document}